# Normal and anti Meyer-Neldel rule in conductivity of highly crystallized undoped microcrystalline silicon films

Sanjay K. Ram[*,a], Satyendra Kumar[*,b] and P. Roca i Cabarrocas[!]

[*]*Department of Physics & Samtel Centre for Display Technologies, Indian Institute of Technology Kanpur, Kanpur-208016, India*

[!]*LPICM, UMR 7647 - CNRS - Ecole Polytechnique, 91128 Palaiseau Cedex, France*

We have studied the electrical conductivity behavior of highly crystallized undoped hydrogenated microcrystalline silicon ($\mu$c-Si:H) films having different microstructures. The dark conductivity is seen to follow Meyer Neldel rule (MNR) in some films and anti MNR in others, which has been explained on the basis of variation in the film microstructure and the corresponding changes in the effective density of states distributions. A band tail transport and statistical shift of Fermi level are used to explain the origin of MNR as well as anti-MNR in our samples. The observation of MNR and anti MNR in electrical transport behavior of $\mu$c-Si:H is discussed in terms of the basic underlying physics of their origin and the significance of these relationships.

## 1. INTRODUCTION

Meyer-Neldel Rule (MNR) [1] is a well-known phenomenon seen in many thermally activated processes, including electronic conduction in amorphous or disordered semiconductors, where it correlates exponentially the conductivity prefactor ($\sigma_0$) and the conductivity activation energy ($E_a$) with the equation:

$$\sigma_0 = \sigma_{00} e^{GE_a} \qquad (1)$$

where $G$ and $\sigma_{00}$ are called MN parameters. Often $G^{-1}$ is denoted as $E_{MN}$, the Meyer-Neldel characteristic energy. The origin of MNR in amorphous silicon ($a$-Si:H) has been a subject of debate, although the model describing the statistical shift of Fermi level ($E_f$) with temperature has been the most accepted model [2,3,4,5].

While $a$-Si:H is a homogeneous material, hydrogenated microcrystalline silicon ($\mu$c-Si:H) is a heterogeneous material consisting of a crystalline phase comprised of crystallite grains, which conglomerate to form columns growing perpendicular to the substrate, and amorphous (or disordered) phase and voids populating the inter-grain and inter-columnar boundaries [6,7]. MNR has been reported in *doped* $\mu$c-Si:H films with MNR parameters similar to those obtained in $a$-Si:H [8]. This was explained in terms of statistical shift model analogous to $a$-Si:H. While the electrical transport properties of mixed-phase $\mu$c-Si:H materials are influenced by the constituent $a$-Si:H phase, the electrical transport in single-phase $\mu$c-Si:H having complete crystallization from the beginning of film growth cannot be compared to the transport in $a$-Si:H [9,10]. In such a material, the absence of an amorphous phase gives rise to mechanisms and routes of electrical transport different from our conventional understanding of relationship between electrical transport and variation in crystallinity [9]

Apart from MNR, another interesting and important phenomenon is the anti-MNR, in which a negative value of $E_{MN}$ is seen. Anti MNR has been reported in heavily doped $\mu$c-Si:H [8,11,12] and heterogeneous Si (*het*-Si) thin film transistors (TFTs) [13]. This phenomenon has been explained by the $E_f$ moving deep into the band tail. The understanding of MNR and anti-MNR in the electrical transport behavior in the context of the inhomogeneous $\mu$c-Si:H microstructure can provide important insight into the electronic transport mechanisms. In this paper, we present the results of dark conductivity ($\sigma_d$) measurements (above room temperature) conducted on a large microstructural range of well-characterized single-phase $\mu$c-Si:H samples having a broad range of microstructural features. Our results show that both MNR and anti-MNR can be observed in $\mu$c-Si:H, depending on the film microstructure.

## 2. EXPERIMENT

The undoped $\mu$c-Si:H films were deposited in a parallel-plate glow discharge plasma enhanced chemical va-por deposition system operating at a standard rf frequency of 13.56 MHz, using high purity $SiF_4$, Ar and $H_2$ as feed gases. Different microstructural series of samples were created by systematically varying gas flow ratios ($R = SiF_4/H_2$) or $T_s$ (100–250°C) for samples having different thicknesses ($d$ ~50–1200nm). We employed bifacial Raman scattering (RS), spectroscopic ellipsometry (SE), X-ray diffraction (XRD), and atomic force microscopy (AFM) for structural investigations. Coplanar $\sigma_d(T)$ measurements were carried out from 300K to 450K on these well-characterized annealed samples having a variety of film thicknesses and microstructures [14].

---

[a] Corresponding author. E-mail address: skram@iitk.ac.in, sanjayk.ram@gmail.com
[b] satyen@iitk.ac.in

## 3. RESULTS

We first briefly describe the results of the microstructural studies carried out on our material. A high total crystalline volume fraction of all the samples was demonstrated by RS and SE. Bruggeman effective medium approximation (BEMA) was used to fit experimental SE data, using published dielectric functions for low-pressure chemical vapor deposited polysilicon with large (pc-Si-l) and fine (pc-Si-f) grains as reference. In the simulation for fitting the experimental SE data, we have used a three layered structure model consisting of a bottom layer interfacing with substrate, a bulk layer and a top surface roughness layer [14].

A high crystalline volume fraction (>90%) in the bulk of the material from the initial stages of growth is demonstrated by SE results, with the rest being density deficit. With film growth, there is improvement in material density and void fraction becomes nil. An amorphous phase is absent from the bulk of the films. SE results revealed the presence of crystallite grains of two distinct sizes throughout the film thickness, although there is a significant variation in the percentage volume fraction of the constituent large ($F_{cl}$) and small grains ($F_{cf}$) with film growth. The SE results were found to be corroborative with the RS results, the deconvolution of which was done using a bimodal size distribution of large crystallite grains (LG ~70–80nm) and small crystallite grains (SG ~6–8nm). With film growth the ratio $F_{cl}/F_{cf}$ rises. The commencement and evolution of conglomeration of crystallites with film growth demonstrated by AFM corresponds to the appearance and increase in the LG fraction. XRD results substantiated the presence of two different sized crystallites and revealed presence of preferred crystalline orientation in (400) and (220) directions fully grown films [14].

Coming to the electrical transport results, at above room temperature, $\sigma_d(T)$ of all the $\mu$c-Si:H films having different microstructures, prepared under different deposition conditions, follows Arrhenius type thermally activated behavior:

$$\sigma_d = \sigma_0 e^{-E_a/kT} \quad (2)$$

However, films having similar thicknesses but deposited under different conditions show disparate electrical transport behaviors, due to the effects of deposition conditions on the underlying film microstructure. The correspondence between $d$ or deposition parameters ($R$ or $T_s$) and film microstructure is not always systematic because the same microstructure can be achieved by the adjustment of one or more deposition parameters. Our earlier work has shown that $F_{cl}$ values correlate well to the stage of film growth and morphology, regardless of the deposition parameters [14]. Our microstructural and correlative electrical transport studies suggest that in single-phase $\mu$c-Si:H material, in the absence of any correspondence with the total crystalline fraction (which is high and un-varying), or amorphous phase (which is non-existent), the electrical transport behavior can be well correlated empirically to the film microstructure with the help of the percentage fraction of the constituent large crystallite grains [10].

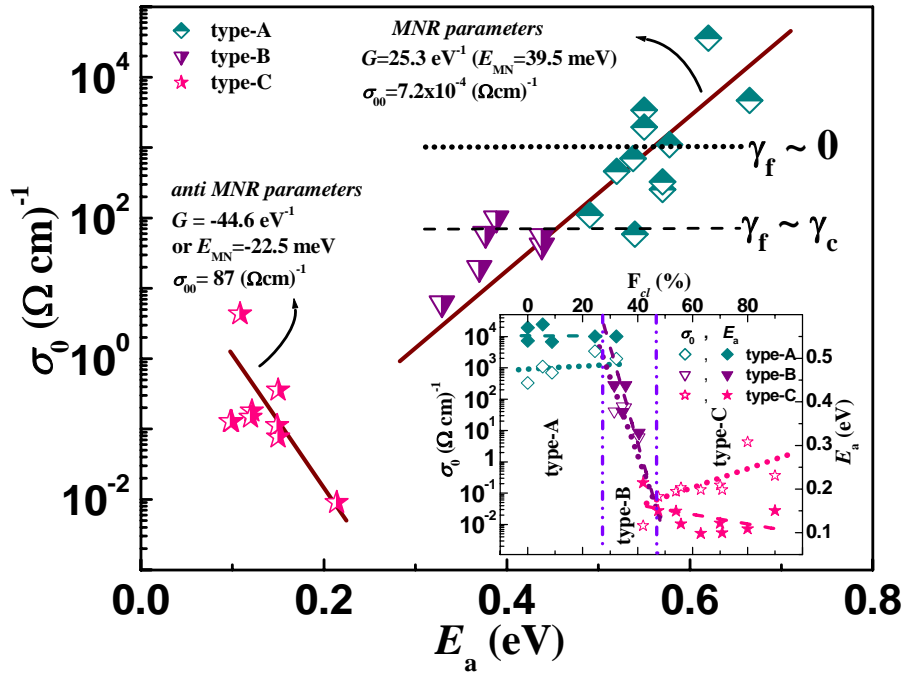

FIG. 1. Correlation between $\sigma_0$ and $E_a$ in *undoped* $\mu$c-Si:H samples (*types*: *A*, *B* and *C*). The samples of *types-A* and *B* follow MNR while *type-C* material shows anti MNR behavior with parameters as shown in the graph. In the MNR region the dotted line indicates the possible position of $\sigma_0$ where $\gamma_f \approx 0$ and the dashed line where $\gamma_f \approx \gamma_c$. The inset shows the classification of highly crystallized $\mu$c-Si:H samples based on the variation of $\sigma_0$ and $E_a$ of $\mu$c-Si:H samples (*types*: *A*, *B* and *C*) with $F_{cl}$.

A systematic study and data analysis of the large number and variety of samples in this study necessitates organizing them into some classes having common attributes. Therefore, on the basis of structural investigations of the $\mu$c-Si:H films at various stages of growth and under different growth conditions, we have segregated out the unique features of microstructure and growth type present in the varieties of films, with respect to the correlative coplanar electrical transport properties and classified them into three *types*: *A*, *B* and *C*. This has been shown concisely in the inset of Fig. 1.

Outlining this classification briefly, the *type-A* films have small grains, low amount of conglomeration (without column formation), and high density of inter-grain boundary (GB) regions containing disordered phase. Here, $F_{cl}$ <30%, $\sigma_0$ and $E_a$ are constant [~$10^3$ $(\Omega cm)^{-1}$ and ~0.55 eV respectively]. The *type-B* films contain a fixed ratio of mixed grains in the bulk. There is a marked morphological variation in these films due to the commencement of conglomeration of grains resulting in column formation, and a moderate amount of disordered phase is present in the columnar boundaries. Here $F_{cl}$ varies from 30% to 45% and there is a sharp drop in $\sigma_0$ [from ~$10^3$ to 0.1 $(\Omega cm)^{-1}$] and $E_a$ (from ~0.55 to 0.2 eV). The *type-C* $\mu$c-Si:H material is fully crystallized, crystallite conglomerates are densely packed with significant fraction of large crystallites (>50%) and preferential orientation is seen. Here $\sigma_0$ shows a rising trend [from 0.05 to 1 $(\Omega cm)^{-1}$] and the fall in $E_a$ is slowed down (from 0.2 to 0.10 eV). The microstructural features that result in such changes in the electrical transport behavior are discussed later.

While a relation between the microstructural parameter $F_{cl}$ and electrical transport parameters is evident in the inset of Fig. 1, there also appears to be some relation between $\sigma_0$ and $E_a$. Therefore, it would be useful to study the variation of $\sigma_0$ with $E_a$ for each type of material. Figure 1 shows a semi-logarithmic plot between $\sigma_0$ and the $E_a$ obtained on our samples. The data for *types-A* and *B* are found to fall along the MNR line. We found the values of the MNR parameters, $G \approx 25.3$ eV$^{-1}$ (or $E_{MN} \approx 39.5$ meV) and $\sigma_{00} \approx 7.2 \times 10^{-4}$ $(\Omega cm)^{-1}$ from the fit shown in the figure. In contrast, the data for samples of *type-C* shows an inverse linear relationship between logarithmic value of $\sigma_0$ and $E_a$. The correlation between $\sigma_0$ and $E_a$ appears to change sign in this case, demonstrating *anti MNR*. The values of anti MNR parameters were determined to be $G \approx -4.6$ eV$^{-1}$ or $E_{MN} \approx -22.5$ meV and $\sigma_{00} \approx 86.8$ $(\Omega cm)^{-1}$.

## 4. DISCUSSION

The observation of MNR in $\mu$c-Si:H material has been discussed in literature, where as anti MNR in undoped $\mu$c-Si:H material has not been reported earlier. Therefore, the phenomena of MNR and anti MNR in electrical transport behavior of undoped highly crystallized $\mu$c-Si:H draws our attention towards the basic physics, in terms of both the origin and significance of these relationships.

In a disordered semiconductor, the density of states (DOS) distribution may not be symmetrical with respect to the band center due to tailing of localized states at the band edges as well as defect states in the gap. Therefore, the experimentally obtained $\sigma_0$ contains terms arising from the two effects. The first comes from the statistical shift of $E_f$ and the second involves a temperature dependent shift of the band edges, i.e., of conduction and valence band edges, $E_c$ and $E_v$ [5]. According to Mott, one can express the conductivity expression as: [15]

$$\sigma_d(T) = \sigma_M \exp(-(E_c - E_f)/kT)) \qquad (3)$$

where $\sigma_M$ is minimum metallic conductivity. $E_c$ and $E_f$ are both dependent on temperature. Approximating the temperature shift of $E_c$ and $E_f$ to be linear functions with the slopes $\gamma_c$ and $\gamma_f$ respectively, we get

$$E_c(T) = E_c^0 - \gamma_c T \quad \text{and} \quad E_f(T) = E_f^0 - \gamma_f T \qquad (4)$$

where $E_c^0$, $E_f^0$ are the positions of $E_c$ and $E_f$ at $T$=0K. After inserting Eq. (4) into Eq.(3), we get Eq.(2) with

$$E_a = E_c^0 - E_f^0 \qquad (5)$$

$$\sigma_0 = \sigma_M \exp\left[(\gamma_c - \gamma_f)/k\right] \qquad (6)$$

The band shifts are taken relative to midgap. They are positive when $E_c$ and $E_f$ move towards midgap.

MNR in $\mu$c-Si:H has been generally understood using the above theory/ calculations, but the anti MNR behavior is rather less elucidated. Most workers [12,13] have attributed the anti MNR behavior observed in doped $\mu$c-Si:H material to the model implicating energy band (EB) diagram of crystalline silicon (c-Si) and a-Si:H interface as proposed by Lucovsky and Overhof (LO) [8]. According to this model, anti MNR can be observed only in a degenerate case when very heavy doping of the $\mu$c-Si:H material causes $E_f$ to move above $E_c$ in the crystalline phase and consequently $E_f$ can move deeply into the tail states in the disordered region. In explaining anti MNR behavior on the basis of this EB diagram [8], equal band edge discontinuities at both ends of c-Si and a-Si:H interfaces were assumed in a study of intrinsic het-Si

TFT [13]. However, the calculation of EB diagram of the interface is a complicated task as the band edge discontinuities have also been debated [16,17].

The linear relation between $\log\sigma_0$ and $E_a$ can be obtained only if $E_f$ lies in the CB tail or close to the minimum of DOS. However, there may exist a combination of factors, which can serve to either augment or diminish the anti MNR effect (or the loss of linearity). For instance, when $E_f$ approaches the boundaries, if there is a flat DOS spectrum near the edge, the statistical shift of $E_f$ will diminish along with a decrease in the temperature derivative of $E_f$. This may cause an occurrence of anti MNR behavior for small and large $E_a$ regions [3,4]. However, such a deviation (anti MNR effect) may diminish if there is a jump in the DOS profile present at the edge of the steep tail [4]. Another situation where the anti MNR is more pronounced at values of $E_a$ on the lower side, is when the DOS value at the minimum reduces [4]. Such a reduction in DOS value at the minimum, where exponential CB and negatively charged dangling bond ($DB^-$) tails meet, has been observed in the case of $n$-type doping of $a$-Si:H [18]. Here DOS of $DB^-$ band increases with increased doping level. The failure of MNR at higher $E_a$ side can be seen when $E_f$ lies far in the tail of the $DB^-$ states or in intrinsic materials where DOS at the mid gap is almost flat, due to which the temperature derivative of $E_f$ will be very little or almost zero [4].

Besides the DOS features, the actual carrier transport route in the material is vital for understanding electronic transport. The above discussed theories/models hold true for $\mu$c-Si:H only if a band tail transport exists. In heavily doped $\mu$c-Si:H material current route follows through crystallites/ columns and transport properties can be understood by well established grain boundary trapping (GBT) models. However, recent experimental evidence suggests a dominant role of the disordered silicon tissue of the boundaries encapsulating the crystallite columns in electrical transport in fully crystalline single phase undoped $\mu$c-Si:H material [9]. In poorly crystalline $\mu$c-Si:H material, where an interconnected network of boundary tissue has not formed, transport is considered to take place through the amorphous matrix [9]. Our material is somewhat different, as it achieves full crystallization from the beginning of the growth, which is especially relevant in the *type-A* material, where there is almost complete crystallization, without an amorphous matrix, though the column formation has not started [14].

Let us now view the observed electrical transport behavior of our materials in the context of their microstructural features and analyze the applicability of the concepts discussed above. In *type-A* material, the question of formation of potential barrier (i.e., transport through crystallites) does not arise because the large number of defect/trap sites (in GBs) compared to free electrons and small size of crystallites will result in a depletion width that is sufficiently large to become greater than the grain size, causing the entire grain to be depleted. Therefore, the transport will be governed by the band tail transport. Corroboratively, the inset of Fig. 1 shows that in *type-A* material, $E_a$ becomes nearly saturated (~0.55eV) and $\sigma_o$ reaches ~$10^3$ $(\Omega cm)^{-1}$. This means the $E_f$ is lying in the gap where the DOS does not vary much and there is a minimal movement of $E_f$, or $\gamma_f \approx 0$ [2,3]. We have indicated this possible position of $\sigma_0$ where $\gamma_f \approx 0$ in Fig. 1 by a dotted line. The initial data points shown in Fig. 1 for *type-A* have higher $\sigma_o$ [~ $10^4$ $(\Omega cm)^{-1}$] and $E_a$ (~0.66eV), because of a shift in $E_c$ and/or a negative value of $\gamma_f$, as happens in $a$-Si:H for $E_a$ towards the higher side [2,3].

*Type-B* material is a crossover region from *type-A* to *type-C* that shows large variations in $\sigma_0$ and its $E_a$ as seen in the inset of Fig. 1. The improvement in film microstructure leads to a delocalization of the tail states causing the $E_f$ to move towards the band edges, closer to the current path at $E_c$. The statistical shift $\gamma_f$, depends on the temperature and the initial position of $E_f$, and when the $E_f$ is closer to any of the tail states and the tail states are steep, $\gamma_f$ is rapid and marked. In Fig. 1, the transition between *type-A* and *type-B* materials shows a few data points somewhat scattered around the MNR line, belonging to both the types, which show a more or less constant $\sigma_o$ [70-90 $(\Omega cm)^{-1}$] with the fall in $E_a$ (0.54-0.40 eV), indicating that the temperature shift of $E_f$ and that of the CB have become equal, canceling each other out (i.e., $\gamma_f \approx \gamma_c$) [2,3]. We have depicted the possible position of $\sigma_0$ where such a situation can occur in Fig. 1 by a dashed line. In this case, the $E_f$ is pinned near the minimum of the DOS between the exponential CBT and the tail of the defect states ($DB^-$) [2,3]. With increasing crystallinity and/or improvement in the microstructure, the minimum shifts towards $E_c$ leading to a decrease of $E_a$.

In *type-C* material, the calculated values of free electron concentrations (from $\sigma_d$, $E_a$ data and mobility $\mu_{TRMC}$ from time resolved microwave conductivity) do not suggest the possibility of unintentional doping achieving such a high value of background doping concentration as to result in degeneracy. Also, in a degenerate case, the conductivity behavior of polycrystalline material is found to exhibit a $T^2$ dependence of $\sigma_d$ [19], which is not the case in our material. Therefore, the EB model as suggested by LO seems inapplicable to our undoped $\mu$c-Si:H case, though the value of $E_{MN}$ here is close to the value reported in heavily doped $\mu$c-Si:H (-20meV) [8]. In *type-C* $\mu$c-Si:H material, a higher $F_{cl}$ and large size of columns (>300nm) result in less columnar boundaries, and therefore less defects associated with the boundaries. Moreover, a conducting network of such interconnected boundaries is formed, resulting in higher conductivity (rise in $\sigma_0$). Transport through this encapsulating disordered tissue requires a band tail transport. The large columnar microstructure results in a long range ordering which is sufficient to delocalize an appreciable range of states in the tail state distribution. In addition, higher density of available free carriers and low value of defect density can cause a large increase in $DB^-$ density together with a decrease in positively charged dangling bond ($DB^+$) states in the gap, which results in a lower DOS near the CB edge and can create a possibility of a steeper

CB tail. In this situation, if $E_f$ is lying in the plateau region of the DOS, it may create an anti MNR situation.

## 5. CONCLUSION

In conclusion, our electrical transport studies show that both MNR and anti MNR can be seen in the dark conductivity behavior of highly crystalline single-phase undoped $\mu$c-Si:H material, depending on the microstructure and the correlative DOS features. Our study strongly indicates the presence of a band tail transport in $\mu$c-Si:H. We have shown that the statistical shift model can successfully explain both the MNR and anti MNR behavior in our material.

## ACKNOWLEDGEMENTS

One of the authors (SKR) gratefully acknowledges Dean of R & D, I.I.T. Kanpur, Samtel Centre for Display Technologies, I.I.T. Kanpur, Council of Scientific and Industrial Research, New Delhi and Department of Science and Technology, New Delhi, for providing financial support.